\documentclass[journal]{IEEEtran}

\title{ \Huge Rate-splitting in the presence of multiple receivers}
\author{
	\IEEEauthorblockN{
	Omar Fawzi
	and  
	Ivan Savov \\ %
	} 
	\IEEEauthorblockA{ %
		School of Computer Science, McGill University,  \textit{Montr\'eal, Qu\'ebec, Canada} \\
	} 
	\vspace{-7mm}
}

\usepackage{amsthm}
\usepackage{amsmath}
\usepackage{amssymb}

\newtheorem{example}{Example}

\def\mcal{\mathcal}

\newcommand{\cC}{\mathcal{C}}

\def\myleq{<}
\def\Sasoglu{\c{S}a\c{s}o\u{g}lu}

\usepackage{tikz}
\usetikzlibrary{arrows,shapes,decorations,automata,backgrounds,petri}
\usetikzlibrary{shapes.gates.logic.US}
\usepackage[latin1]{inputenc}
\usetikzlibrary{calc,patterns,decorations.pathmorphing,decorations.markings}

\begin{document}
\maketitle

\begin{abstract}
In the presence of multiple senders, one of the simplest decoding strategies 
that can be employed by a receiver is \emph{successive decoding}.
In a successive decoding strategy, 
the receiver decodes the messages one at a time using the knowledge of the previously decoded messages as side information.
Recently, there have been two separate attempts to construct codes for the interference channel using successive decoding based on the idea of \emph{rate-splitting}.

In this note, we highlight a difficulty that arises when a rate-splitting codebook is to be decoded by multiple receivers. 
The main issue is that the rates of the split codebook are tightly coupled to the properties of the channel to the receiver,
thus, rates chosen for one of the receivers may not be decodable for the other.
We illustrate this issue by scrutinizing two recent arguments claiming to achieve the Han-Kobayashi rate region for the interference channel using rate-splitting and successive decoding.

\end{abstract}

\section{Introduction} 

The interference channel describes an information transmission 
scenario where two communication links are forced to share the communication medium. 
In the discrete memoryless setting, we use a conditional probability
distribution $p(y_1,y_2|x_1,x_2)$ to model both the interaction between the
transmitted signals as well as the noise that is inherent to the channel.
The generality of the interference channel model and the fact that there are multiple senders
and multiple receivers involved, makes it difficult to find a coding strategy
that is optimal.

In this note, we are concerned with the rates that are achievable using a simple decoding strategy that we call \emph{successive simple decoding}. A simple decoder can only distinguish between \emph{independent} codewords. A successive simple decoder may use many simple decoders one at a time given the knowledge of the previously decoded messages as side information. 
Such strategies are called \emph{simple} because of their lower decoding complexity, which 
would make them easier to implement in practice. Indeed, the complexity of the successive
decoding strategy is no different from the point-to-point decoder \cite{urbanke-rate-splitting}.
Another motivation for studying such decoders is in multi-user \emph{quantum} information theory, where constructions of \emph{simultaneous} decoders is only known for some special channels; see \cite{FHSSW11}. A typical example of a decoder that is not simple is a \emph{simultaneous} decoder sometimes also called \emph{jointly-typical} decoder. 
For the interference channel, it is thus a natural question to ask whether there is a successive simple decoding strategy to achieve the well  known Han-Kobayashi rate-region? Recently, two separate conference papers have appeared that claim to answer this question in the affirmative using different variants of rate-splitting \cite{sasoglu2008successive,yagi2011multi}. In this paper, we show that these arguments are incomplete. More generally, we point out that a codebook obtained using rate-splitting depends on the intended receiver and that such a coding scheme should be analyzed with great care when there are multiple %
receivers.

This paper is organized as follows. In Section \ref{sec:rate-splitting-mrx}, we give a brief introduction to rate-splitting in the simplest case of a point-to-point communication scenario and show with a specific example that a rate-splitting codebook can in general only be decoded by the receiver for which the rate-splitting was designed. In Section \ref{sec:rate-splitting-ic}, we analyse more precisely codes for the interference channel by considering the arguments of \cite{sasoglu2008successive} and \cite{yagi2011multi}.

\section{Rate-splitting for multiple receivers}
\label{sec:rate-splitting-mrx}

Consider a point-to-point discrete memoryless channel described by some transitions probabilities denoted $p(y|x)$ for $x \in \mcal{X}$ and $y \in \mcal{Y}$. When coding for $n$-uses of such a channel, the most natural random codebook construction is defined by $\cC = \{x^n(m)\}, m \in [1:2^{nR}]$ where the codewords $x^n(m)$ are drawn randomly from $\prod^n p(x)$ and \emph{independently} for each message. 
It is then standard to show that provided $R < I(X;Y)$, a receiver obtaining the outcome of the channel can decode the message with small error probability. Observe here that if we have two channels $p(y|x)$ and $p(y'|x)$, then \emph{both} receivers can decode the \emph{same} codebook $\cC$ as long as $R < I(X;Y)$ and $R < I(X; Y')$. In particular, for any channel $p(y'|x)$ satisfying $I(X;Y') \geq I(X;Y)$ (i.e., $Y'$ carries more information about $X$ than $Y$ does), the codebook $\cC$ is decodable from the output of the channel $p(y'|x)$.

In more complicated scenarios involving multiple users, it is sometimes useful to consider codebooks obtained using rate-splitting. Some reasons why such a strategy might be useful will become clear in Section \ref{sec:rate-splitting-ic}, but this is not important for the discussion in this section; see \cite{urbanke-rate-splitting} for a more detailed treatment of rate-splitting. A rate-splitting codebook is obtained by picking two probability distributions $p(x_a)$ and $p(x_b)$ on the input alphabet $\mcal{X}$ and a mixing function $f : \mcal{X} \times \mcal{X} \to \mcal{X}$. The rate-splitting codebook is then defined as $\cC_{\rm rs} = \{f^n(x^n_{a} (m_{a}), x^n_{b} (m_{b})) \},  m_{a} \in [1: 2^{n R_{a}}], m_{b} \in [1: 2^{n R_{b}}]$, where the codewords $\{ x^n_{a}(m_{a}) \}$ and $\{ x^n_{b}(m_{b}) \}$ are chosen randomly and independently according to $p^n(x^n_a)$ and $p^n(x^n_b)$. The overall rate of $\cC_{\rm rs}$ is $R=R_a + R_b$. Observe that the codewords of $\cC_{\rm rs}$ are \emph{not} independent in general. 

For a given channel $p(y|x)$, under what conditions on $R_a$ and $R_b$ is the codebook $\cC_{\rm rs}$ decodable for the receiver? Assuming we want a simple decoding strategy, then the receiver has no choice but to either decode $m_a$ first then $m_b$ or vice-versa. The first case is possible if and only if $R_a < I(X_a; Y)$ and $R_b < I(X_b; Y|X_a)$ and the second if and only if $R_b < I(X_b; Y)$ and $R_a < I(X_a; Y|X_b)$. Suppose we fix $R_a$ and $R_b$ satisfying these conditions. Now consider another channel  $p(y'|x)$, can we make a similar statement that $\cC_{\rm rs}$ is decodable from the output of $p(y'|x)$ as long as $I(X;Y') \geq I(X; Y)$? The main objective of this note is to point out this is \emph{not true}. In fact, the condition $R_a < I(X_a; Y)$ and $R_b < I(X_b; Y|X_a)$ heavily depends on the inner workings of the channel $p(y|x)$ rather than just the mutual information between the input and the output. We illustrate this in the following example.

\begin{example}
\label{ex:split-no-good}
Consider a situation with a single sender connected to two different
receivers by point-to-point communication channels.
The first channel is described by $(\mcal{X}=\{0,1,2,3\}, p_1(y_1|x),\mcal{Y}_1=\{0,1,2,3\})$, 
where $y_1$ is the output of Receiver~1 and
the probability distribution is the ``broken typewriter channel''
from $\{0,1,2,3\}$ to $\{0,1,2,3\}$, i.e.,  $p_1(i|i)=p_1(i+1 \!\mod 4\ |\ i)=\frac{1}{2}$.
The second channel is the ``most-significant-bit channel'' from $\{0,1,2,3\}$ to 
$\{0, 1\}$, with transition probabilities $p_2$ such that
$p_2(0|0)=p_2(0|1)=1$ and $p_2(1|2)=p_2(1|3)=1$.

Let $X$ be uniformly distributed on $\mcal{X}$. Then we have $I(X; Y_1) = I(X; Y_2) = 1$, so by using the standard codebook $\cC$ with $R < 1$, both receivers can decode the message with small error probability.

Consider now the case where we prepare a certain rate-splitting codebook and we pick the rates at the convenience of Receiver~1. We chose $\epsilon = \frac{1}{2}$ in the $\min$ construction described in \cite{urbanke-rate-splitting}. The construction gives the following entropic quantities: $I(X_{a}; Y_1) > 0.270838, I(X_{b};Y_2|X_{a}) > 0.729161$. This means that we can choose the rate-splitting codebook $\cC_{\rm rs}$ with $R_a = 0.270838 $ and $R_b = 0.729161$, and it will be decodable by Receiver~1. Now we turn to Receiver~2 for whom the entropic quantities are as follows: $I(X_{a}; Y_2) < 0.311279$, $I(X_{b};Y_2|X_{a}) < 0.688722$ and $I(X_{b};Y_2) < 0.459148$, $I(X_{a};Y_2|X_b) < 0.540853$. In other words, if Receiver~2 tries to decode $m_a$ first then $m_b$, he will succeed in decoding $m_a$ but then fail in decoding $m_b$ correctly as $R_b > I(X_{b}; Y_2|X_a)$. In addition, if Receiver~2 tries to decode $m_b$ first, he will fail because $R_b > I(X_b; Y_2)$. Even more, Receiver~2 can simply not decode the code $\cC_{\rm rs}$ regardless of the strategy he uses (e.g., even if he uses a jointly-typical decoder). This is because the rate $R_b$ is simply too large to transmit on the channel defined by $p'(y_2,x_a|x_b) = p(x_a) p_2(y_2|f(x_a, x_b))$.
\end{example}

\section{Rate-splitting for the interference channel}
\label{sec:rate-splitting-ic}

	Rate-splitting techniques were originally proposed in the
	context of the multiple access channel (MAC) 	\cite{urbanke-rate-splitting,rimoldi2001generalized}.
	It is shown that by splitting the messages of the senders
	and choosing the rates of the split codebook appropriately,
	it is possible for the receiver to use successive simple decoding 
	in order to achieve all the rates in the MAC capacity region.

	The interference channel with two senders and two receivers 
	$p(y_1,y_2|x_1,x_2)$ induces two MAC sub-channels one for each receiver:
	$p(y_1|x_1,x_2)$ and $p(y_2|x_1,x_2)$.
	One possible coding strategy for the interference channel is to 
	build a codebook for each multiple access channel that is decodable for 
	\emph{both} receivers.
	Two recent papers propose the use of rate-splitting strategies for the 
	interference channel \cite{sasoglu2008successive, yagi2011multi}.	
	In this section, we analyze these to proposals and point out how 
	the remarks from Section~\ref{sec:rate-splitting-mrx} apply to the coding
	strategies proposed in these papers.
	We will show that the arguments in these papers are incomplete,
	and that in general finding a 
	rate-splitting  strategy for the interference channel faces significant obstacles
	when some messages are to be decoded by both receivers.	

	The receivers are not required to decode messages which are not intended for them,
	but doing so can lead to better rates.
	The capacity-achieving coding strategies for the special case of 
	interference channels with \emph{strong} and \emph{very strong} interference 
	both require the decoding of the interfering messages \cite{carleial1975case,costa1987capacity}.
	A more general strategy is that of Han and Kobayashi which 
	involves \emph{partial} decoding of the interfering messages \cite{HK81}. 

	In the remainder of this section, we will discuss coding strategies 
	which require each receiver to decode the messages of both senders: $m_1$ and $m_2$.
	Note that the discussion also applies to the Han-Kobayashi strategy
	where we identify $m_1$  as the ``common'' message of Sender~1
	and $m_2$ with the common message of Sender~2.

	\subsection{\Sasoglu \ strategy}		

		The rate-splitting construction for the MAC with two senders 
		contains a freedom about which sender's message to split.
		One approach would be to split the message of Sender~1 $m_1= (m_{1a},m_{1b})$ 
		and the receiver would decode in the order $m_{1a} \to m_2 \to m_{1b}$.
		Another approach would be to split the messages of Sender~2 
		$m_2 = (m_{2a},m_{2b})$  and in this case the receiver will decode
		 in the order $m_{2a} \to m_1 \to m_{2b}$.
		Since there are also two receivers, 		\Sasoglu \ observed,
		it would be possible to choose the split of one message
		for Receiver~1 and the split of the other message for Receiver~2 \cite{sasoglu2008successive}.

		\Sasoglu \ chooses to split each codebook at the convenience of the \emph{other} receiver.
		Thus, Sender~1 will use a codebook 
		$\cC_{1\rm rs} = \{f_1(x^n_{1a} (m_{1a}), x^n_{1b} (m_{1b})) \},
		m_{1a} \in [1: 2^{n R_{1a}}],
		m_{1b} \in [1: 2^{n R_{1b}}]$,
		and choose the rates $R_{1a}$ and $R_{1b}$ so that Receiver~2 can
		decode in the order $m_{1a} \to m_2 \to m_{1b}$:
		\begin{align}
		 R_{1a}  &\myleq  I(X_{1a};Y_2), \label{r1a-dec-cond} \\
		 R_{2}     &\myleq  I(X_{2};Y_2|X_{1a}),  \label{rtwo-dec-cond}\\ 
		 R_{1b}  &\myleq  I(X_{2b} ; Y_2 | X_{1a} X_{2} ). \label{r1b-dec-cond}
		\end{align}%
		Similarly, Sender~2 will use a rate-splitting codebook $\cC_{2\rm rs}$ in which 
		we split the message $m_{2} = (m_{2a}, m_{2b})$ 
		and the rates are chosen as required for the MAC sub-channel to Receiver~1
		and the decode ordering $m_{2a} \to m_1 \to m_{2b}$.
		
		Thus, in order to account for both receivers, Sender~1 will use $\cC_{1\rm rs}$ and Sender~2
		will use $\cC_{2\rm rs}$. It might seem at first that we have just shown that as long as the rates
		satisfy \eqref{r1a-dec-cond}-\eqref{r1b-dec-cond} and the analogous	inequalities obtained by
		substituting `$1$' and `$2$', the rate $(R_{1a} + R_{1b}, R_{2a} + R_{2b})$ is achievable
		for the interference channel using a successive simple decoding strategy. However, a closer
		inspection reveals that we run into the issue raised in Section~\ref{sec:rate-splitting-mrx}. In fact,
		when analyzing the decoding of Receiver~2, we assumed that the codebook used by
		Sender~1 is a standard codebook with independent codewords but we later replaced it
		with a rate-splitting codebook $\cC_{1 \rm rs}$. 
		A simple successive decoding strategy for Receiver~2 would require
		decoding the message $m_{2}$ in parts,
		as in the ordering $m_{1a} \to m_{2a}\to m_{2b} \to m_{1b}$,
		which  is only possible if the rates $R_{2a}  \myleq  I(X_{2a};Y_2),
		 R_{2b}  \myleq  I(X_{2b} ; Y_2 | X_{1a} )$,
		 which, as seen in Example~\ref{ex:split-no-good}, will not hold in general despite the fact that the 
		 overall rate $R_{2}$ satisfies \eqref{rtwo-dec-cond}.

\begin{figure}[htbp]
		\centering
		\includegraphics[width=2.4in]{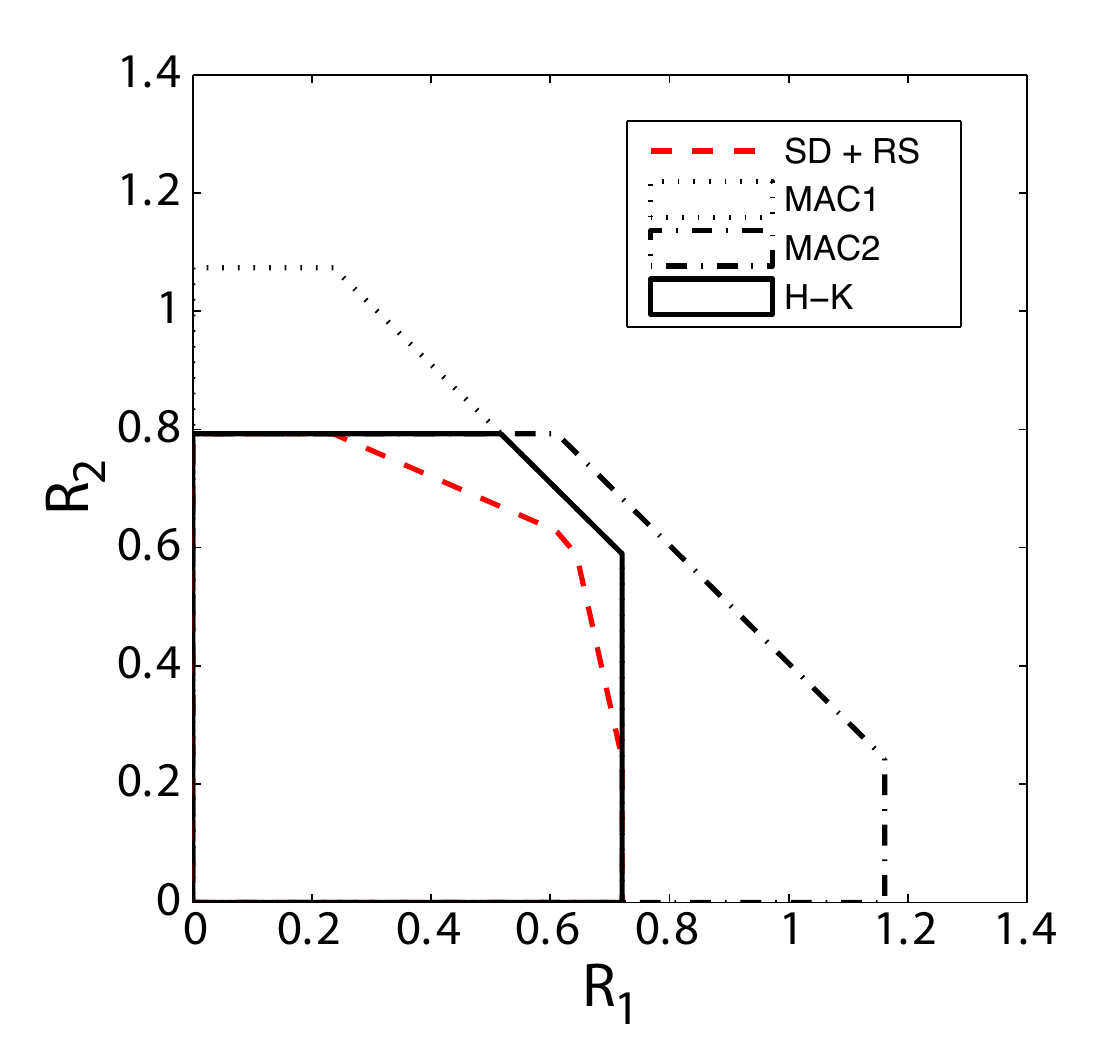}
		\vspace{-3mm}
		
		\caption{	Comparison of the rates achievable by successive decoding and rate-splitting
				 (successive simple decoding) and the simultaneous decoding used by
				 the Han-Kobayashi strategy in the context of a Gaussian interference channel.
				The input power of both senders is $P_1=P_2=2$,
				and the noise powers are $N_1=0.35$ and $N_2=0.3$.
				The channel coefficients are $g_{11} = \sqrt{0.3}$,
				$g_{22} = \sqrt{0.3}$, $g_{12} = \sqrt{0.6}$ and  $g_{21} = \sqrt{0.6}$,
				which means that this channel exhibits strong interference.
				Observe that the SD+RS strategy is suboptimal. %
		}
		\label{fig:sdrs-vs-hk}		
		\end{figure}

		We conclude that to strictly analyze the rate region that is achievable
		via this rate-splitting strategy we need to take into account the decoding abilities
		of both receivers and choose the rates for the codebooks as follows:
		\begin{align*} 
		 R_{1a}  &\leq  \min\{  I(X_{1a};Y_2), %
		 I(X_{1a};Y_1|X_{2a}) %
		  \}, \\
		 R_{1b}  &\leq  \min\{ I(X_{1b} ; Y_2 | X_{1a} X_{2} ),   %
						    I(X_{1b};Y_1|X_{2a}X_{1a})
		    \}, \\
		 R_{2a}  &\leq  \min\{  I(X_{2a};Y_1), 
						 I(X_{2a};Y_2|X_{1a})
		 \}, \\
		 R_{2b}  &\leq  \min\{ I(X_{2b} ; Y_1 | X_{2a}X_{1} ), 	%
						    I(X_{2b};Y_2|X_{1a}X_{2a})
		    \}, 
		\end{align*}
		therefore the region achievable by this strategy is in general sub-optimal as illustrated in Figure \ref{fig:sdrs-vs-hk}.

	\vspace{-7mm}

	\subsection{Switch-based rate-splitting strategy} %

		Recently, in work independent from \cite{sasoglu2008successive}, 
		a rate-splitting strategy for the interference channel was proposed in \cite{yagi2011multi}.
		Yagi and Poor describe a coding strategy in which 
		the message of Sender~2 is split using the \emph{generalized time sharing} approach~\cite{rimoldi2001generalized}.
		The message of Sender~2 is split into \emph{four} parts: 
		
		{\small
		\begin{equation}
		m_{2}
		 =
		\left[
		\begin{array}{cc}
		m_{2a} & m_{2b}  \\[1mm]
		m_{2c} & m_{2d}  
		\end{array}
		\right].
		\end{equation}}%
		Each of the split messages $m_{2\alpha}, \; \alpha \in [a,b,c,d]$ is encoded
		in a random codebook $\{ x^{n}_{2\alpha} (m_{2\alpha})\}$, ${ m_{2\alpha} \in [1:2^{nR_{2\alpha} }] }$,
		each codeword being generated randomly and independently according to
		the probability distribution for Sender~2 $p^n(x^n_2)$.
		The two codebooks  $\{ x^{n}_{2a}(m_{2a})\}$ and $\{ x^{n}_{2b}(m_{2b})\}$ are then combined
		to form the codebook $\{ x^{n}_{2ab}(m_{2a},m_{2b}) )\}$ based on $n$ instances of
		a switch random variable $S_{h} \in \{\ell,r\}$ with $x_{2abi}=x_{2ai}$ if $S_{hi}=\ell$,
		and $x_{2abi}=x_{2bi}$ if $S_{hi}=r$.
		A codebook $\{ x^{n}_{2cd}(m_{2c},m_{2d}) )\}$ is  similarly constructed
		for the bottom row. 
		The codebooks $\{ x^{n}_{2ab}(m_{2a},m_{2b}) )\}$
		and $\{ x^{n}_{2cd}(m_{2c},m_{2d}) )\}$ are then combined using
		another switch random variable $S_v \in \{t,b\}$ 
		to form the codebook $\{ x^{n}_{2}(m_{2a},m_{2b},m_{2c},m_{2d}) )\}$
		used by Sender~2.
		The codebook for Sender~1 is not split: $x^n_1(m_1)$, $m_1 \in [1:2^{nR_1} ]$.
		Receiver~1 will decode the messages in the order
		$(m_{2a},m_{2b}) \to m_{1} \to  (m_{2c},m_{2d})$,
		while Receiver~2 will  decode in the order $(m_{2a},m_{2c}) \to m_{1} \to  (m_{2b},m_{2d})$.
		The rates $R_{2\alpha}$ %
		can be chosen such that the {\bf h}orizontal split 
		of the rates is adapted for the decoding of Receiver~1, 
		while the {\bf v}ertical split 
		is adapted for Receiver~2.

		Let us consider the first step in the decoding performed by 
		Receiver~1 more closely.
		Receiver~1 is required to decode the codebook $\{ x^{n}_{2ab}(m_{2a},m_{2b}) )\}$,
		which contains the two messages $m_{2a}$ and $m_{2b}$.
		Observe, however,  that the codewords of the codebook are not independent.
		This means that choosing the overall rate to be
		$R_{2ab} = R_{2a}+R_{2b} \leq I(X_{2ab};Y_1)$ 
		is not a sufficient condition to conclude the codebook is decodable.
		In order to show that Receiver~1 can perform the decoding,
		we must ensure that each of the constituent messages can be decoded.
		The receivers can use either simultaneously decoding for $m_{2a}$ and $m_{2b}$
		or use  successive simple decoding: $m_{2a}  \to m_{2b}$.
		In both cases, the decoding of $m_{2b}$ is only possible if 
		$R_{2b}   \myleq  I(X_{2b}; Y_1|X_{2a})$,
		which will not be the case in general because the rate $R_{2b}$ was chosen 
		at the convenience of Receiver~2.

\vspace{-1mm}
\section{Conclusion}
	 	\label{sec:conclusion}

Rate-splitting is an important coding strategy for multi-user communication scenarios that consists in adding ``structure'' to a standard random codebook construction which allows the message to be decoded in parts. Having such a structure means that the different codewords of a rate-splitting codebook are not independently chosen. We showed here that this comes at a cost: such a codebook is not as easily decodable as a standard codebook whose codewords are chosen independently.

We should stress that we did not prove that successive simple decoding strategies are strictly weaker than general decoding strategies. Rather, we showed that the strategies that were previously proposed have overlooked an issue that arises when using rate-splitting in the presence of multiple receivers. We leave it as an open problem to determine whether successive simple decoding strategies can be used to achieve the Han-Kobayashi rate region for the interference channel.
Proving that this is possible would probably require new techniques or at least many layers of rate-splitting.

\section*{Acknowledgements}
We would like to thank Patrick Hayden, Eren \c{S}a\c{s}o\u{g}lu, Pranab Sen, Mai Vu, Mark Wilde, Hideki Yagi for helpful discussions. This research was supported by CIFAR, NSERC and ONR grant No. N000140811249.

\bibliographystyle{IEEEtran}
\bibliography{interferenceChannel}

\end{document}